\documentclass[preprint,amsmath,amssymb,aps,prb,floatfix]{revtex4-1}
\usepackage{graphicx,dcolumn,bm,txfonts}

\newcommand*{\et}[0]{\textit{et~al.}}
\newcommand{\Fig}[1]{Fig.~\ref{fig:#1}}
\newcommand{\Tab}[1]{Table~\protect\ref{tab:#1}}

\newcommand*{\eps}[0]{\varepsilon}

\newcommand*{\be}[0]{\begin{equation}}
\newcommand*{\ee}[0]{\end{equation}}

\begin{document}
\title{First-principles calculation of H vibrational excitations at
a dislocation core of Pd}
\author{Hadley M. Lawler}
\author{Dallas R. Trinkle}
\email{dtrinkle@illinois.edu}
\affiliation{Department of Materials Science and Engineering, University of
Illinois, Urbana-Champaign}
\date{\today}

\begin{abstract}
Palladium is an ideal system for understanding the behavior of hydrogen in
metals.  In Pd, H is located both in octahedral sites and in dislocation
cores, which act as nanoscale H traps and form Cottrell atmospheres.
Adjacent to a dislocation core, H experiences the largest possible
distortion in $\alpha$-Pd.  Ab initio density-functional theory computes
the potential energy for a hydrogen in an octahedral site in $\alpha$-Pd
and in a trap site at the core of a partial of an edge dislocation.  The Pd
partial dislocation core changes the environment for H, distorting the H-Pd
bonding which changes the local potential, vibrational spectra, and
inelastic form factor for an isolated H atom.  The decrease in excitation
energy is consistent with experiments, and the calculations predict
distortions to the H wavefunction.
\end{abstract}
\pacs{}

\maketitle

Renewable energy requires new methods for the production, storage and
transportation of energy from the point of production.  The potential for
hydrogen as an energy storage medium\cite{Schlapbach2001} has renewed
interest in the fundamentals of hydrogen in metals---a topic with a long
history.\cite{Myers} In particular, the ease of catalysis of molecular to
atomic hydrogen on the surface of palladium has motivated the study of
atomic hydrogen and hydride formation in Pd.  Hydrogen's vibrational
excitations in metals provide an interesting window into H behavior due to
its low mass; and $\alpha$-Pd is a useful system to consider as a model
system that is simple to prepare.\cite{PdH} There are various means of
measuring the vibrational energies, such as conductance
spectroscopy,\cite{cond} Raman scattering,\cite{Raman} and inelastic
neutron scattering.\cite{Drexel} Neutron scattering measured isotopic
effects,\cite{Antonov} linewidths and their dependence on
temperature,\cite{Chowdhury} the optical band and its high-energy
features,\cite{Rowe73} and concentration dependence of the
spectra.\cite{Rowe84} Heuser \et\ recently measured the hydrogen excitation
peak at 4K to be 59meV for the dilute (0.08at.\%) H concentration, while at
300K the peak was 68meV.\cite{Heuser08} The low concentration suggests
hydrogen trapped at edge-dislocation core interstitial sites.  This is due
to the strong interactions of dislocations and hydrogen,\cite{Kirchheim}
with a binding energy of 0.2eV.\cite{Heuser98} To determine the effect of
the dislocation core on hydrogen vibration spectra, we compute the
anharmonic H potential energy adjacent to a dislocation core from
first-principles, and calculate excitation energies and wavefunctions.  The
result is a decrease in excitation energy consistent with experiments and
distortions in the hydrogen wavefunctions due to strain and symmetry
breaking.

The anharmonic potential for hydrogen in Pd increases vibrational
excitation energies and produces anisotropy in the inelastic form factor
compared with a harmonic potential.\cite{Elsasser,Kemali} The low mass of H
produces zero-point motion of 0.15\AA\ in an octahedral interstitial
environment in $\alpha$-Pd, which in turn samples a potential energy
surface that is no longer purely harmonic.  This increases transition
energies for H vibration in Pd by $\sim 50\%$ relative to the harmonic
approximation.\cite{Elsasser} The vibrational spectrum for H has the
degeneracy of an octahedral potential,\cite{Krimmel} and excitations which
are not simple multiples of the first excitation energy.\cite{Rowe84} The
wavefunctions have cubic symmetry, and variations in the inelastic form
factor with scattering wavevector show modulations between high symmetry
directions.\cite{Hempelmann,Kemali} However, when a hydrogen atom occupies
an octahedral site near a dislocation core, the cubic symmetry is
completely broken, and H experiences a distorted environment.

Predicting the vibrational excitations for a hydrogen atom requires
accurate computation of the potential energy for hydrogen.
Density-functional theory calculations are performed with
\textsc{vasp}\cite{Kresse93,Kresse96b} using a plane-wave basis
with the projector augmented-wave (PAW) method,\cite{Blochl1994} with
potentials generated by Kresse.\cite{Kresse1999} The local-density
approximation as parametrized by Perdew and Zunger\cite{Perdew1981} and a
plane-wave kinetic-energy cutoff of 250eV ensures accurate treatment of the
potentials.  The PAW potential for Pd treats the $s$- and $d$-states as
valence, and the H $s$-state as valence.  The restoring forces for H in Pd
change by only 5\%\ compared with a generalized gradient approximation, or
including Pd $4p$-states in the valence; our choice of the local-density
approximation is computationally efficient, and gives an $\alpha$-Pd
lattice constant of 3.8528\AA\ compared with the experimentally measured
3.8718\AA.  To compute the dynamical matrix for the lattice Green
function\cite{TrinkleLGF2008}, and to relax H at the octahedral site in
$\alpha$-Pd, we use a $4\times4\times4$ simple-cubic supercell of 256
atoms, with a $6\times6\times6$ k-point mesh; while the dislocation
geometry with 382 atoms uses a $1\times1\times8$ k-point mesh.  The
electron states are occupied using a Methfessel-Paxton smearing of 0.25eV.
For the H octahedral site in $\alpha$-Pd and the partial dislocation core,
atom positions are relaxed using conjugate gradient until the forces are
less than 5meV/\AA.

\begin{figure}[htb]
\includegraphics[width=2.5in]{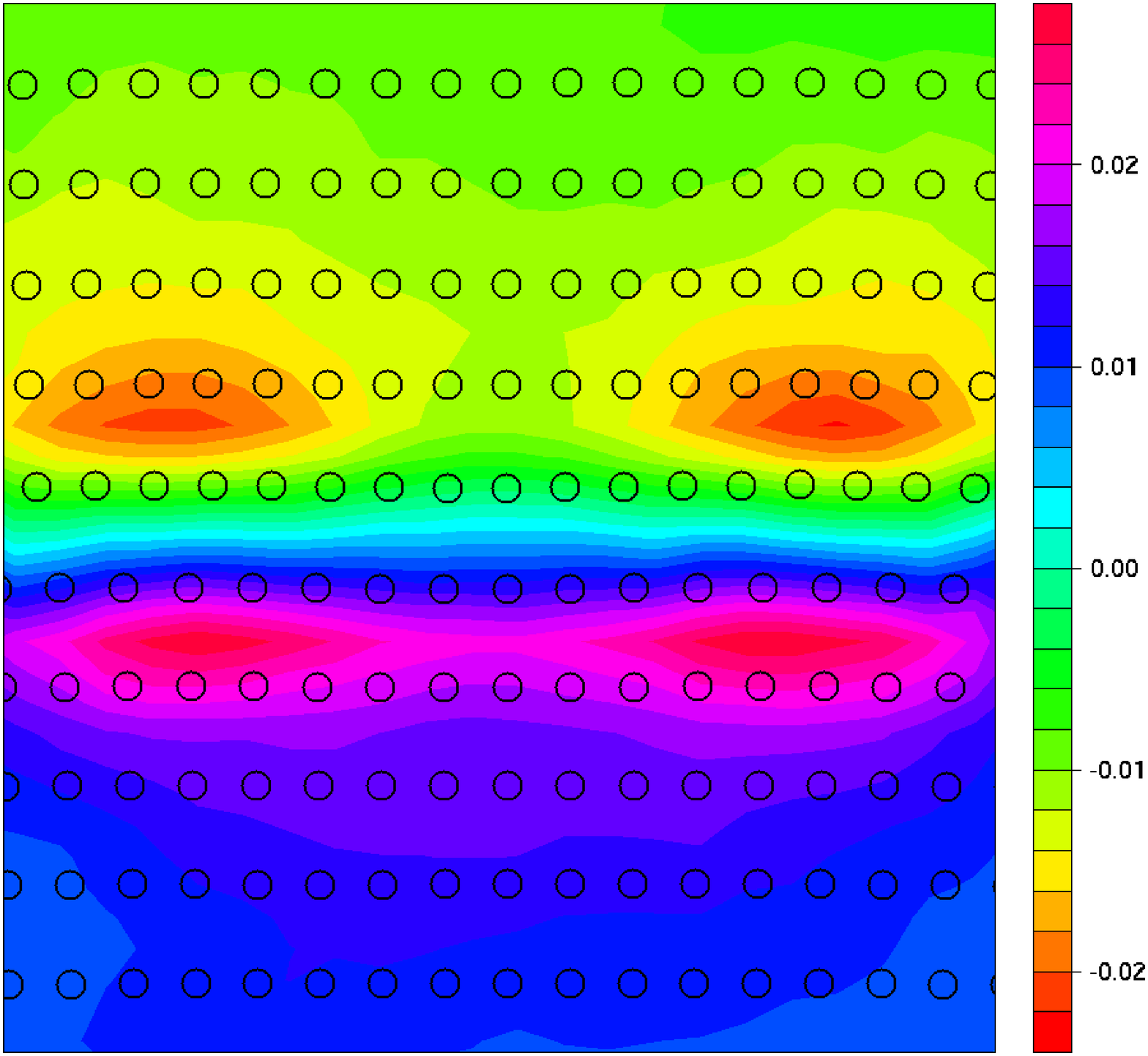}\\
\includegraphics[width=2.5in]{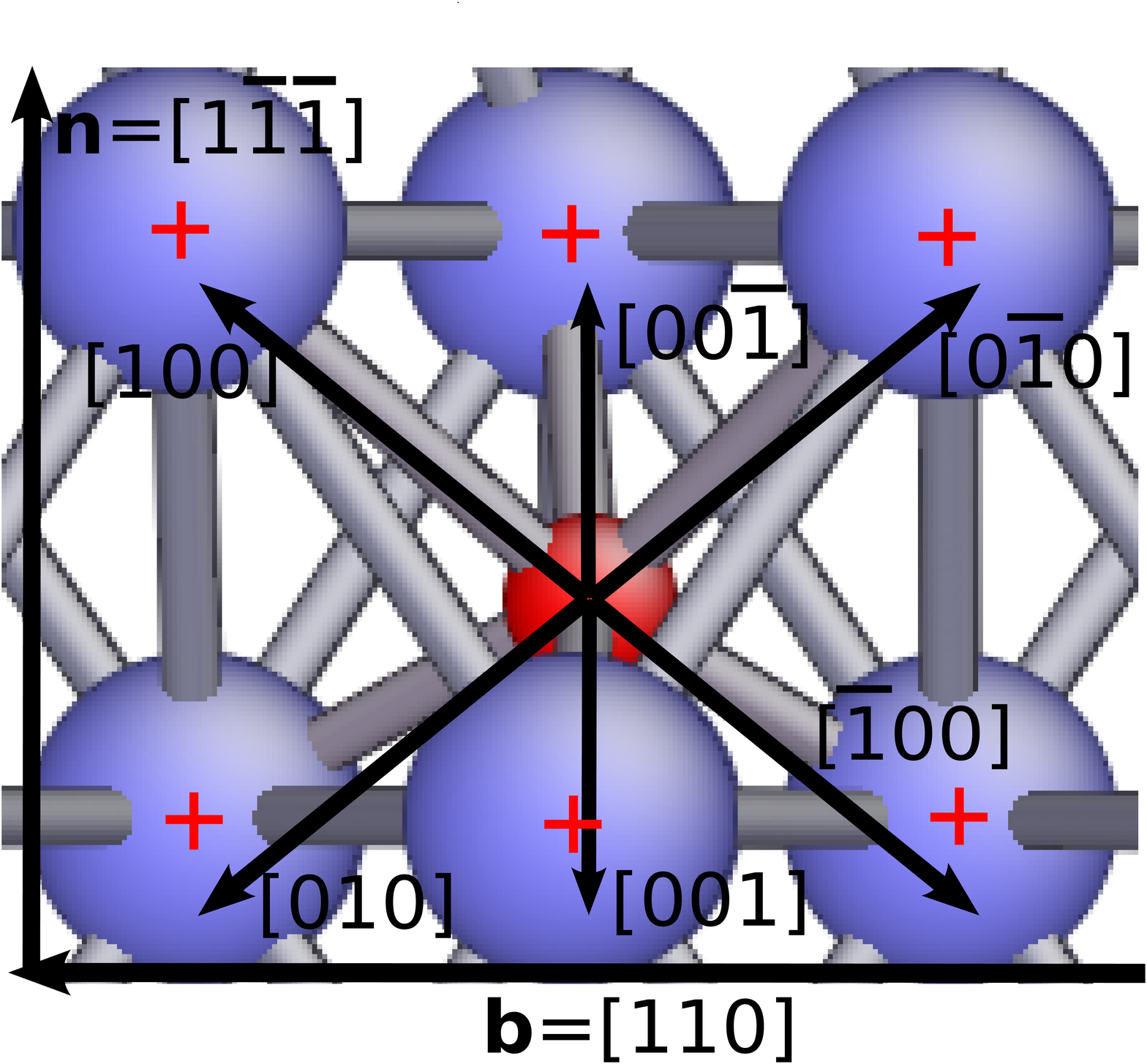}
\caption{(a) Strain at octahedral sites in Pd edge-dislocation core; (b)
Hydrogen (red) in the octahedral site below the Pd (blue) partial
dislocation core in the $(1\bar1\bar1)$ slip-plane.  The edge dislocation
has a total Burgers vector $\frac{a}{2}[110]$, while the partials have
Burgers vector $\frac{a}{6}[211]$ and $\frac{a}{6}[12\bar1]$ and are
separated by $6.5b$. (b) The hydrogen is in an octahedral site with large
tensile strain; the top three Pd atoms are the most expanded, and the
bottom three are less expanded.  The dislocation core displaces the
neighboring Pd atoms for the octahedral site from the relaxed bulk
positions shown by the $\langle 100\rangle$ arrows.  Moreover, the hydrogen
atom moves away from the top three atoms ($[100]$, $[00\bar1]$, and
$[0\bar10]$) increasing the distance to 1.96\AA--2.16\AA.  The hydrogen
moves closer to the other three atoms, decreasing the distance to
1.87\AA--1.91\AA, compared to a distance of 1.96\AA\ for H in an octahedral
site in fcc-Pd.}
\label{fig:geometry} 
\end{figure}

First-principles calculations using lattice Green function-based flexible
boundary conditions compute a stress-free edge dislocation in Pd.  Flexible
boundary conditions embed a dislocation in an infinite bulk which responds
harmonically to forces.\cite{Sinclair1978,Rao1998} The harmonic response is
captured by the lattice Green function, which gives the displacement field
necessary to relax a line-force in an infinite harmonic crystal; it is
computed directly from the force-constant matrix.\cite{TrinkleLGF2008} The
initial geometry is a periodic ``slab'' supercell of 382 fcc Pd atoms: an
infinite cylinder with free surfaces along the dislocation threading line
$\frac{a}{2}[1\bar12]$ in a $9a[110]\times 11\frac{a}{2}[\bar111]$ box.
Anisotropic elasticity gives the initial displacements for an
$\frac{a}{2}[110]$ edge dislocation centered in the cylinder.  Flexible
boundary conditions use forces from density-functional theory, and relaxes
each atom based on its location within one of three different regions
determined by the distance from the dislocation
core.\cite{Sinclair1978,Rao1998} The 45 atoms around the core start with
non-zero forces, and conjugate-gradient minimization relaxes them while the
other atoms are fixed.  The neighboring 100 atoms between the core and the
free surfaces act as a coupling region to a virtual infinite bulk.  They
start with zero forces, but displacing the core atoms induces forces.  With
the lattice Green function, we can displace all 382 Pd atoms to relax the
forces in the coupling region.  This includes the outer 237 atoms that have
non-zero forces due to their proximity to the free surfaces; their forces
are ignored in the calculation.  The relaxation cycle continues alternating
between conjugate gradient near the core and lattice Green function
relaxation in the intermediate region, until all forces are less that
5meV/\AA.  The final result is the stress-free dislocation core equilibrium
geometry: an edge dislocation split into two $\frac{a}{6}\langle
211\rangle$-type partials separated by a distance of $6.5b$
(c.f.~\Fig{geometry}).\cite{Heuser2009}

\Fig{geometry} shows the distorted local environment for a hydrogen atom
adjacent to a partial dislocation core.  Interstitial atoms favor regions
of expansion, and below the slip plane of the partial dislocation core has
maximum tensile strain and non-volumetric effects, such as differences in
the bondlengths along the slip-plane normal.  Considering the plane
$(1\bar1\bar1)$ halfway between the three Pd atoms above and below, the H
atom is displaced away from the partial core (above the top three atoms).
The three Pd atoms above the interstitial site have longer Pd-H bondlengths
than the three below it.  For an idealized edge-dislocation, the slip-plane
normal points along the strain gradient from compressive to tensile, and
just below the slip plane the predominant strain is uniaxial tension along
the Burgers vector.  Changes in the bondlength above and below the
interstitial site are greatest for the [100] and [010] octahedral axes,
which are non-orthogonal to the Burgers vector.  This effect is such that
although there is an overall $5\%$ volume expansion about the interstitial
site, the [100] and [010] octahedral axes expand by an average of $3\%$,
and the [001] axis contracts by $2\%$.  For the octahedral axes
non-orthogonal to the Burgers vector, there are expansion and contraction
of Pd-H bondlengths above and below the interstitial site, while
bondlengths along the [001] are approximately the same.

The potential energy for H in an octahedral site in $\alpha$-Pd and a
partial dislocation core are fit to force versus displacement data for the
hydrogen atom from its relaxed position.  Displacements of 0.025\AA,
0.15\AA\ and 0.2\AA\ provide data of restoring force on H, where the
0.025\AA\ displacements give the harmonic limit and the larger
displacements provides information about the anharmonic terms.  For
$\alpha$-Pd and $\alpha$-Pd with a 5\%\ expansion in volume, all small
displacement directions are equivalent; we use $[100]$, $[110]$, and
$[111]$ directions for 0.2\AA\ displacements for a total of four
displacements.  For the partial dislocation core, cubic symmetry is broken:
we use all six $\langle 100\rangle$ small displacements and six $\langle
100\rangle$, twelve $\langle 110\rangle$, and eight $\langle 111\rangle$
displacements for a total of 58 displacements.  The potential is
represented as a fourth order polynomial in displacements $x_i$; excluding
linear and constant terms, this gives 31 terms to be fit (reducing to 3 in
the case of cubic symmetry).  This representation of the potential
reproduces the first-principles forces to less than 1meV/\AA.%
\footnote{See EPAPS Document No. E-PRBPDH-XX-XXXXXXX for restoring
force data and polynomial potential coefficients.  This document may be
found in the online article's HTML reference section, via the EPAPS
homepage (http://www.aip.org/pubservs/epaps.html), or from ftp.aip.org in
the directory /epaps/.  See the EPAPS homepage for more information.}

\begin{figure*}[ht]
\includegraphics[width=\textwidth]{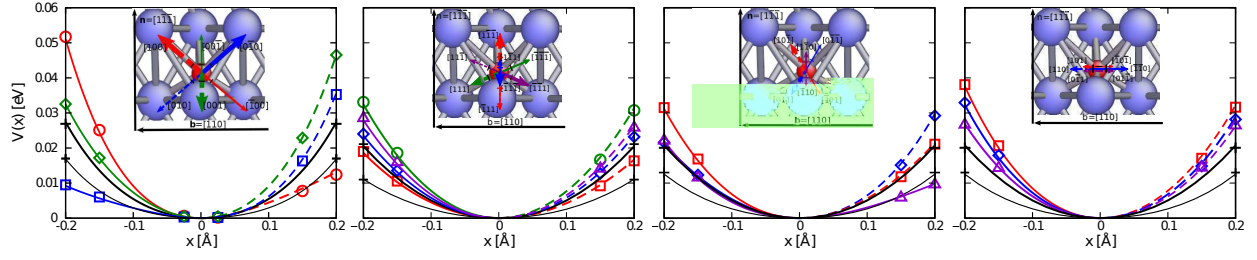}
\caption{
Hydrogen potential energy as a function of displacement along (a) $\langle
100\rangle$, (b) $\langle 111\rangle$, (c) $\langle 110\rangle$ directions
in the slip plane, and (d) $\langle 110\rangle$ directions out of the slip
plane.  Thick black shows response in $\alpha$-Pd, and thin black shows
response in 5\% expanded $\alpha$-Pd.  Positive (negative) displacements
along corresponding directions are dashed (solid).  Arrows show
corresponding displacements, for the potential curves of the same color and
linestyle.  Thick displacement arrows have positive out-of-page
components; negative for thin arrows.  Single points on the plots are the
results of individual first-principles calculations, and the continuous
curves are fits to the Hellmann-Feynman forces.  (a) The cube directions
$[100]$ (circles), $[010]$ (squares), and $[001]$ (diamonds) show the
greatest change to the potential due to the dislocation core.  The softest
directions, $[100]$ and $[0\bar10]$, point towards regions of expansion,
while the $[\bar100]$ and $[010]$ stiff directions bring the hydrogen
closer to Pd atoms.  (b) The octahedral directions $[111]$ (circles),
$[\bar111]$ (squares), $[1\bar11]$ (diamonds), and $[11\bar1]$ (triangles)
show less change due to the dislocation core.  The most significant
softening is for the octahedral direction corresponding to the slip plane
normal. (c) The out-of-plane closed-packed directions $[10\bar1]$
(squares), $[011]$ (diamonds), and $[\bar110]$ (triangles) are similarly
softened in the direction of expansion and stiffened in the opposite
direction compared with bulk. (d) The in-plane closed-packed directions
$[101]$ (squares), $[110]$ (diamonds), and $[0\bar11]$ (triangles) are all
stiffened compared with bulk.}
\label{fig:potential} 
\end{figure*}

\Fig{potential} shows that potentials along directions of the shortened 
bondlengths are steeper than $\alpha$-Pd, while those along elongated
bondlengths are shallower.  For an octahedral site in $\alpha$-Pd, the
$\langle100\rangle$ directions are the stiffest, while the
$\langle111\rangle$ and $\langle110\rangle$ directions are shallower.  In
addition, the $\langle110\rangle$ directions connect neighboring octahedral
sites and $\langle111\rangle$ pass through an octahedral face.  All of the
potentials soften with the expansion of the lattice by 5\%.  The softest of
the $\langle100\rangle$-type directions have components along the Burgers
vector, and are above the interstitial site, i.e. [100] and [0$\bar{1}$0].
The resultant potential for coordinates along these directions has large
cubic terms, reflecting bondlength variation above and below the
interstitial site.  The potential for the [001] coordinate is less cubic,
consistent with the smaller change in bondlength along this direction.  The
distortions to the potential from the dislocation core geometry is
reflected in the broken symmetry for excitations and the resulting
wavefunctions for the hydrogen atom.

\Tab{transition} shows that the predicted transition energies from our
potentials; it is an overestimate compared to experimentally measured peaks
in scattering intensity, but the reduction in transition energy matches
well.  To solve for the energy spectrum of our potential, we use a basis of
products of Hermite polynomials in the three displacement directions $x_i$
with a fundamental length $x_0$.  The fundamental length is chosen to
harmonically match the anharmonic potential: the harmonic ground state
$|000\rangle$ is the solution to a 1D harmonic potential with stiffness
$\langle 000| \nabla^2 V |000\rangle/3$.  For $\alpha$-Pd,
$x_0=0.215\text{\AA}$ and for a partial dislocation core
$x_0=0.210\text{\AA}$; note that these are larger than the anharmonic
ground-state RMS spread.  The full basis $|m_1m_2m_3\rangle$ contains all
$m_i$ up to $m_1+m_2+m_3\le M$, for a total of $(M+3)(M+2)(M+1)/6$ basis
functions.  Energies for the first three excited states are converged to
0.3meV with $M=8$ (165 basis functions) and 0.03meV with $M=16$ (969 basis
functions).  Our overestimation of the transition energies is consistent
with other density-functional theory calculations for the octahedral site
in $\alpha$-Pd;\cite{Ke} it is also independent of the exchange-correlation
potential and treatment of the ionic cores.  The reduction in excitation
energy of 7meV in the dislocation core corresponds with the experimental
measurement of 10meV.

Cubic symmetry is broken at the partial dislocation core, which splits
the degeneracy of the excited state, and gives three principal axes for the
ground state wavefunction spread.  The eigenvectors of the spatial
covariance matrix, $X^{(2)}_{ij}=\langle x_ix_j\rangle-\langle
x_i\rangle\langle x_j\rangle$, give the directions of maximum and minimum
spread.  The ground-state eigenvectors are $\vec{n}_1=[0.66,-0.75,-0.06]$,
$\vec{n}_2=[0.73,0.66,-0.16]$, and $\vec{n}_3=[0.16,0.06,0.99]$---roughly,
$[1\bar10]$, $[110]$, and $[001]$---with RMS values of 0.19\AA, 0.16\AA,
and 0.14\AA; compared to 0.15\AA\ for H in $\alpha$-Pd.  The soft direction
$\vec n_1 \approx [1\bar10]$ corresponds with the softest of the nearly
symmetric direction in the potential (c.f. \Fig{potential}c, triangles);
the stiffest direction $\vec n_3 \approx [001]$ corresponds with the
stiffest nearly symmetric direction in the potential
(c.f. \Fig{potential}a, diamonds), and $\vec n_2 \approx [110]$ a mutually
orthogonal ``bulk-like'' potential (c.f. \Fig{potential}d, diamonds).
Moreover, the three directions correspond to the shapes of the first three
excited states in \Tab{transition}.

The first three excited states are $p$-like, and have maximum spreads along
directions which correspond to $\vec{n}_1$, $\vec{n}_2$, and $\vec{n}_3$.
The first excited state has a maximum spread along $[0.68,-0.73,-0.07]$,
the second along $[0.71,0.68,-0.17]$, and the third along
$[0.15,0.11,0.98]$.  The magnitudes of maximum spread for each of the three
states are 0.35\AA, 0.27\AA, and 0.23\AA, compared to a maximum spread of
0.25\AA\ for the first-excited state in $\alpha$-Pd.  The expanded volume
in the partial dislocation core decreases the transition energy; however
\Tab{transition} shows degeneracy-splitting due to broken symmetry from the
local strain.  This strain allows for a low lying transition state $\vec
n_1$ that is 30meV below the similar transition in $\alpha$-Pd, or 20meV
below in the expanded lattice.  This is a direct consequence of the
hydrogen occupying a site near the partial dislocation core.

\begin{figure}[ht]
\includegraphics[width=2.75in]{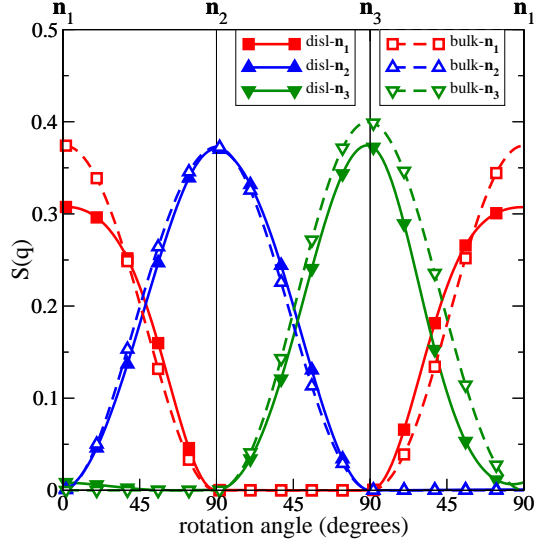}
\caption{Inelastic form factor $S(\vec q)$ for a H adjacent to the
dislocation core and in octahedral bulk sites.  The momentum direction
$\vec q$ rotates between the three orthogonal vectors $\vec n_i$,
corresponding to the directions of maximal $(\vec n_1$) and minimal ($\vec
n_3$) spread of the H ground state wavefunction in a dislocation core.  The
peak along each $\vec n_i$ indicates that the excited state is $p$-like
along that direction.  The calculation of form factors for \textit{bulk}
excited states which maximize overlap with the dislocation core states
shows a similar peak shape with direction.  The height of the peaks
corresponds to the spreading of the excited states along those directions in
comparison with the ground state: increased spreading of the excited states
along the particular direction decreases the peak height.}
\label{fig:Sq} 
\end{figure}

\Fig{Sq} shows the orientation of the first three excited states by plotting the 
inelastic form factor, $S(\vec q)$ from the ground state to each excited
state.  The inelastic form factor, for an excitation from the ground state
$|0\rangle$ to excited state $|\vec n_i\rangle$, $S(\vec q)=
\left|\langle 0|\exp(-i\vec {q} \cdot \vec {r})|\vec{n}_i\rangle
\right|^{2}$ determines the intensity of inelastic scattering with
scattering direction $\vec q$.  In \Fig{Sq}, we rotate $\vec q$ from $\vec
n_1$ to $\vec n_2$ to $\vec n_3$ and back to $\vec n_1$.  For comparison,
we also compute $S(\vec q)$ with bulk excited states projected along $\vec
n_i$.  The excited states have similar structure to the excited states in
$\alpha$-Pd site.  The changes in inelastic form factors are due to the
expansion of the excited state along the $\vec n_i$ directions compared
with bulk, where expanded wavefunctions decrease $S(\vec q)$.
Density-functional theory calculations combined with accurate treatment of
boundary conditions compute the relaxed geometry for a hydrogen atom
adjacent to a partial dislocation core in Pd.  We extract the local
potential energy for the hydrogen atom from first-principles to predict the
quantum mechanical transition energies, and compare to the transition
energies in an octahedral site in $\alpha$-Pd.  The changes in the
excitation energies can be directly traced to the distortions in geometry
around the hydrogen atom, the potential energy, and ultimately the
wavefunctions for hydrogen.  The predicted decrease due to the volumetric
expansion is similar to the change in experimental measurements at low
temperatures where the hydrogen occupancy of partial dislocation cores
should come to dominate.  This provides an important step in understanding
the changes in vibrational spectra from hydrogen in Pd as temperature
changes and the hydrogen atoms migrate to more energetically favorable
sites in the sample.  In addition, the potential provides a starting point
for considering pipe diffusion of H along dislocation cores in Pd.

\begin{acknowledgments}
This research was supported by NSF under grant number DMR-0804810, and in
part by the NSF through TeraGrid resources provided by NCSA and TACC.
\end{acknowledgments}

\newpage

\begin{table}
\caption{Calculated and measured transition energies for 
H in $\alpha$-Pd and adjacent to a partial dislocation core.  Hydrogen
occupies an octahedral site in $\alpha$-Pd; the transition from the ground
state to the triply-degenerate first excited state is larger than the
experimentally measured peak.  A volumetric expansion of 5\% reduces the
predicted transition energy by 9meV.  The partial dislocation core produces
a similar volumetric expansion, but includes additional distortion that
lift the degeneracy of the first excited state, giving a set of three
low-lying transition energies.  The labels $\vec n_i$ correspond to the
different orientations of the excited state wavefunction.}
\label{tab:transition}
\begin{ruledtabular}
\begin{tabular}{ccccc}
 &Theory [meV] &Experiment [meV]\\ 
\hline
$\alpha$-Pd& 87 & 69 [\protect\onlinecite{Rowe84}]\\
$\alpha$-Pd ($\eps_V=5\%$)& 78 &\\
\hline
$\vec n_1$& 56 & \\
$\vec n_2$& 80 & 59 [\protect\onlinecite{Heuser08}]\\
$\vec n_3$& 102 &
\end{tabular}
\end{ruledtabular}
\end{table}

\end{document}


\title{First-principles calculation of H vibrational excitations at
a dislocation core of Pd: Supporting online material}
\author{Hadley M. Lawler}
\author{Dallas R. Trinkle}
\email{dtrinkle@illinois.edu}
\affiliation{Department of Materials Science and Engineering, University of
Illinois, Urbana-Champaign}
\date{\today}

\begin{abstract}
H restoring force versus displacement in $\alpha$-Pd, with 5\%\ volume
expansion, and in partial dislocation core.  Polynomial fit for hydrogen
potential energy in same three environments.
\end{abstract}
\pacs{}

\maketitle

\begin{longtable*}{@{\extracolsep{\fill}}cccc@{\quad}|ccc}
\caption{Pd atomic positions relative to a H atom, and H restoring force
versus displacement at an octahedral site near a partial dislocation core
in Pd.  All distances are in \AA\ and forces in eV/\AA.}\label{tab:coreF}\\

\multicolumn{7}{c}%
{\begin{tabular}{cccc}
\hline\hline
Pd [$x$]	&Pd [$y$]	&Pd [$z$]	&Pd [$r$]\\
\hline
 +2.1562	&--0.1321	&+0.0016	&2.1602\\
--1.8553	&--0.1860	&--0.1090	&1.8678\\
 +0.2754	&+1.8880	&--0.0351	&1.9083\\
 +0.1341	&--2.1461	&--0.0170	&2.1504\\
 +0.1869	&--0.1148	&+1.8790	&1.8918\\
 +0.1788	&--0.1726	&--1.9491	&1.9649\\
\hline
\end{tabular}
}\\

\hline
H [$\delta_x$]	&H [$\delta_y$]	&H [$\delta_z$]	&H [$|\delta|$]	&$F_x$	&$F_y$	&$F_z$\\
\hline
\endfirsthead
\multicolumn{7}{c}%
{\tablename\ \thetable{} (continued)}\\
H [$\delta_x$]	&H [$\delta_y$]	&H [$\delta_z$]	&H [$|\delta|$]	&$F_x$	&$F_y$	&$F_z$\\
\hline
\endhead

\multicolumn{7}{c}{{(continued on next page)}} \\
\endfoot

\hline \hline
\endlastfoot

 +0.0250	&0	&0	&0.0250	&--0.0209	&--0.0035	&--0.0028\\
--0.0250	&0	&0	&0.0250	&+0.0436	&+0.0115	&+0.0064\\
 0	&+0.0250	&0	&0.0250	&--0.0009	&--0.0206	&+0.0021\\
 0	&--0.0250	&0	&0.0250	&+0.0141	&+0.0216	&+0.0010\\
 0	&0	&+0.0250	&0.0250	&+0.0019	&+0.0045	&--0.0389\\
 0	&0	&--0.0250	&0.0250	&+0.0112	&+0.0034	&+0.0384\\
\hline
  +0.1500	&0	&0	&0.1500	&--0.0855	&--0.0289	&--0.0189\\
--0.1500	&0	&0	&0.1500	&+0.4072	&+0.0722	&+0.0430\\
 0	&--0.1500	&0	&0.1500	&+0.0353	&+0.0618	&+0.0004\\
 0	&+0.1500	&0	&0.1500	&--0.0680	&--0.2842	&+0.0089\\
 0	&0	&+0.1500	&0.1500	&--0.0468	&+0.0245	&--0.3651\\
 0	&0	&--0.1500	&0.1500	&+0.0127	&+0.0170	&+0.2447\\
\hline
 0	&+0.1061	&+0.1061	&0.1500	&--0.0677	&--0.1353	&--0.1881\\
 0	&--0.1061	&--0.1061	&0.1500	&+0.0391	&+0.0543	&+0.1692\\
--0.1061	&0	&--0.1061	&0.1500	&+0.2519	&+0.0524	&+0.1548\\
 +0.1061	&0	&+0.1061	&0.1500	&--0.0891	&--0.0091	&--0.2527\\
--0.1061	&--0.1061	&0	&0.1500	&+0.2781	&+0.0719	&+0.0274\\
 +0.1061	&+0.1061	&0	&0.1500	&--0.0993	&--0.2083	&--0.0085\\
\hline
 0	&+0.1061	&--0.1061	&0.1500	&--0.0302	&--0.1488	&+0.1392\\
 0	&--0.1061	&+0.1061	&0.1500	&--0.0032	&+0.0533	&--0.2292\\
--0.1061	&0	&+0.1061	&0.1500	&+0.1834	&+0.0542	&--0.1575\\
 +0.1061	&0	&--0.1061	&0.1500	&--0.0637	&--0.0161	&+0.1592\\
--0.1061	&+0.1061	&0	&0.1500	&+0.16259	&--0.07984	&+0.02941\\
 +0.1061	&--0.1061	&0	&0.1500	&--0.05727	&+0.03962	&--0.01581\\
\hline
 +0.0866	&+0.0866	&+0.0866	&0.1500	&--0.1007	&--0.1365	&--0.1711\\
--0.0866	&--0.0866	&--0.0866	&0.1500	&+0.2233	&+0.0673	&+0.1352\\
--0.0866	&+0.0866	&+0.0866	&0.1500	&+0.0911	&--0.0486	&--0.1086\\
 +0.0866	&--0.0866	&--0.0866	&0.1500	&--0.0428	&+0.0367	&+0.1322\\
\hline
 +0.0866	&--0.0866	&+0.0866	&0.1500	&--0.0658	&+0.0377	&--0.2014\\
--0.0866	&+0.0866	&--0.0866	&0.1500	&+0.1380	&--0.0549	&+0.1176\\
 +0.0866	&+0.0866	&--0.0866	&0.1500	&--0.0796	&--0.1479	&+0.1133\\
--0.0866	&--0.0866	&+0.0866	&0.1500	&+0.1708	&+0.0664	&--0.1310\\
\hline
--0.2000	&0	&0	&0.2000	&+0.6714	&+0.1104	&+0.0657\\
 +0.2000	&0	&0	&0.2000	&--0.1022	&--0.0349	&--0.0233\\
 0	&+0.2000	&0	&0.2000	&--0.1121	&--0.4854	&+0.0136\\
 0	&--0.2000	&0	&0.2000	&+0.0388	&+0.0781	&+0.0012\\
 0	&0	&--0.2000	&0.2000	&+0.0032	&+0.0314	&+0.3769\\
 0	&0	&+0.2000	&0.2000	&--0.0801	&+0.0408	&--0.5928\\
\hline
 0	&--0.1414	&--0.1414	&0.2000	&+0.0409	&+0.0612	&+0.2424\\
 0	&+0.1414	&+0.1414	&0.2000	&--0.1040	&--0.2078	&--0.2711\\
--0.1414	&0	&--0.1414	&0.2000	&+0.3813	&+0.0782	&+0.2023\\
 +0.1414	&0	&+0.1414	&0.2000	&--0.1028	&--0.0059	&--0.3836\\
--0.1414	&--0.1414	&0	&0.2000	&+0.4267	&+0.0791	&+0.0407\\
 +0.1414	&+0.1414	&0	&0.2000	&--0.1174	&--0.3327	&--0.0091\\
\hline
 0	&+0.1414	&--0.1414	&0.2000	&--0.0544	&--0.2320	&+0.1817\\
 0	&--0.1414	&+0.1414	&0.2000	&--0.0173	&+0.0566	&--0.3468\\
 +0.1414	&0	&--0.1414	&0.2000	&--0.0727	&--0.0158	&+0.2308\\
--0.1414	&0	&+0.1414	&0.2000	&+0.2705	&+0.0784	&--0.2222\\
 +0.1414	&--0.1414	&0	&0.2000	&--0.0678	&+0.0472	&--0.0200\\
--0.1414	&+0.1414	&0	&0.2000	&+0.2386	&--0.1188	&+0.0417\\
\hline
--0.1155	&--0.1155	&--0.1155	&0.2000	&+0.3284	&+0.0769	&+0.1775\\
 +0.1155	&+0.1155	&+0.1155	&0.2000	&--0.1211	&--0.2055	&--0.2430\\
 +0.1155	&--0.1155	&--0.1155	&0.2000	&--0.0517	&+0.0435	&+0.1885\\
--0.1155	&+0.1155	&+0.1155	&0.2000	&+0.1240	&--0.0680	&--0.1449\\
\hline
--0.1155	&+0.1155	&--0.1155	&0.2000	&+0.1948	&--0.0769	&+0.1444\\
 +0.1155	&--0.1155	&+0.1155	&0.2000	&--0.0794	&+0.0425	&--0.2982\\
--0.1155	&--0.1155	&+0.1155	&0.2000	&+0.2458	&+0.0737	&--0.1821\\
 +0.1155	&+0.1155	&--0.1155	&0.2000	&--0.0964	&--0.2252	&+0.1497\\
\end{longtable*}

\begin{table}
\caption{Pd atomic positions relative to a H atom, and H restoring force
versus displacement at an octahedral site in $\alpha$-Pd.  All distances
are in \AA\ and forces in eV/\AA.}
\label{tab:alphaF}
\begin{tabular}{cccc}
\hline\hline
Pd [$x$]	&Pd [$y$]	&Pd [$z$]	&Pd [$r$]\\
 +1.9555	&0	&0	&1.9555\\
--1.9555	&0	&0	&1.9555\\
 0	&+1.9555	&0	&1.9555\\
 0	&--1.9555	&0	&1.9555\\
 0	&0	&+1.9555	&1.9555\\
 0	&0	&--1.9555	&1.9555\\
\hline
\end{tabular}
\\
\begin{tabular*}{0.9\textwidth}{@{\extracolsep{\fill}}cccc@{\quad}|ccc}
\hline
H [$\delta_x$]	&H [$\delta_y$]	&H [$\delta_z$]	&H [$|\delta|$]	&$F_x$	&$F_y$	&$F_z$\\
\hline
 +0.0250	&0	&0	&0.0250	&--0.0239	&0	&0\\
 +0.2003	&0	&0	&0.2003	&--0.3489	&0	&0\\
 +0.1418	&+0.1418	&0	&0.2005	&--0.1723	&--0.1723	&0\\
 +0.1157	&+0.1157	&+0.1157	&0.2003	&--0.1217	&--0.1217	&--0.1217\\
 +0.2897	&+0.2897	&+0.2897	&0.5018	&--0.3745	&--0.3745	&--0.3745\\
\hline\hline
\end{tabular*}
\end{table}

\begin{table}
\caption{Pd atomic positions relative to a H atom, and H restoring force
versus displacement at an octahedral site in $\eps_V=5\%$ expanded
$\alpha$-Pd.  All distances are in \AA\ and forces in eV/\AA.}
\label{tab:alpha+5F}
\begin{tabular}{cccc}
\hline\hline
Pd [$x$]	&Pd [$y$]	&Pd [$z$]	&Pd [$r$]\\
\hline
 +1.9849	&0	&0	&1.9849\\
--1.9849	&0	&0	&1.9849\\
 0	&+1.9849	&0	&1.9849\\
 0	&--1.9849	&0	&1.9849\\
 0	&0	&+1.9849	&1.9849\\
 0	&0	&--1.9849	&1.9849\\
\hline
\end{tabular}
\\
\begin{tabular*}{0.9\textwidth}{@{\extracolsep{\fill}}cccc@{\quad}|ccc}
\hline
H [$\delta_x$]	&H [$\delta_y$]	&H [$\delta_z$]	&H [$|\delta|$]	&$F_x$	&$F_y$	&$F_z$\\
\hline
 +0.0254	&0	&0	&0.0254	&--0.0129	&0	&0\\
 +0.2036	&0	&0	&0.2036	&--0.2479	&0	&0\\
 +0.1441	&+0.1441	&0	&0.2038	&--0.1098	&--0.1098	&0\\
 +0.1176	&+0.1176	&+0.1176	&0.2036	&--0.0725	&--0.0725	&--0.0725\\
\hline\hline
\end{tabular*}
\end{table}

\begin{table}
\caption{Polynomial expansion for potential energy for hydrogen fit to
force versus displacement data.}
\label{tab:polynomial}
\be
\begin{split}
V_{\alpha\text{-Pd}}(\vec r) =& 
+0.5004 (x^2 + y^2 + z^2) 
+4.6069 (x^4 + y^4 + z^4) 
-3.7379 (x^2 y^2 + y^2 z^2 + z^2 x^2)
\\
V_{+5\% \alpha\text{-Pd}}(\vec r) =& 
+0.2487 (x^2 + y^2 + z^2) 
+4.3358 (x^4 + y^4 + z^4) 
-2.2241 (x^2 y^2 + y^2 z^2 + z^2 x^2)
\\
V_\text{core}(\vec r) =& 
+0.6397 x^2 +0.4187 y^2 +0.7677 z^2 
+0.3074 x y -0.0248 y z +0.1870 z x
\\
&
-2.3703 x^3 +1.6500 y^3 +0.8837 z^3
\\
&
+0.9825 x y^2 +1.0302 z^2 x -0.8675 x^2 y 
-0.8461 y z^2 -0.4979 z x^2 -0.1783 y^2 z
\\
&
-0.1212 x y z
\\
&
+4.0708 x^4 +3.5534 y^4 +5.5368 z^4
\\
&
-2.4136 x^2 y^2 -3.2983 y^2 z^2 -3.4471 z^2 x^2
\\
&
+1.6832 x y^3 +0.5148 z^3 x +1.3317 x^3 y 
+0.0190 y z^3 +0.8594 z x^3 -0.1617 y^3 z
\\
&
+0.3463 x^2 y z -0.3546 x y^2 z -0.9699 x y z^2
\end{split}
\end{equation*}
\end{table}